\documentclass[aps,prd,twocolumn]{revtex4-2}

\usepackage{amssymb}
\usepackage{amsmath}

\usepackage{graphicx}
\usepackage[export]{adjustbox}

\usepackage{xparse}

\usepackage{hyperref}
\usepackage{hypernat}

\usepackage[caption=false,labelfont=normalsize,labelformat=empty,textfont=normalsize,justification=centering]{subfig}

\renewcommand{\Im}{\mathrm{Im}}

\newcommand{\iu}{\mathrm{i}}
\newcommand{\eq}{\mathrm{eq}}

\begin{document}

\title{Cutting rules on a cylinder: a bottom-up approach to quantum kinetic theory for the early universe}

\author{Tom\'a\v s Bla\v zek}
\email{tomas.blazek@fmph.uniba.sk}
\author{Peter Mat\'ak}
\email{peter.matak@fmph.uniba.sk}
\affiliation{Department of Theoretical Physics, Comenius University,\\ Mlynsk\'a dolina, 84248 Bratislava, Slovak Republic}

\date{\today}

\begin{abstract}
Nonequilibrium quantum field theory is often used to derive an approximation for the evolution of number densities and asymmetries in astroparticle models when a more precise treatment of quantum thermal effects is required. This work presents an alternative framework using the zero-temperature quantum field theory, $S$-matrix unitarity, and classical Boltzmann equation as starting points leading to a set of rules for calculations of thermal corrections to reaction rates. Statistical factors due to on-shell intermediate states are obtained from the cuts of forward diagrams with multiple spectator lines. It turns out that it is equivalent to cutting closed diagrams on a cylindrical surface.
\end{abstract}

\maketitle

\section{Introduction.}

Exploring particle physics models' implications in cosmology, we often focus on their predictions for dark matter relic density or $CP$ asymmetries in the standard model sector. Despite sophisticated modern methods being developed, at low temperatures, the classical Boltzmann equation is often used for the sake of simplicity. The particles' interactions are computed in terms of the zero-temperature quantum theory, while the kinetic description of the thermal medium is entirely classical. 

Improving the calculation's accuracy demands the inclusion of thermal corrections. Though this may be straightforward at leading order in the coupling constants, higher-order calculations necessary for matter asymmetry generation require more advanced treatment to be consistent with no asymmetry produced in thermal equilibrium \cite{Sakharov:1967dj}. The usual (top-down) procedure to derive the quantum-corrected Boltzmann equation is based on the \emph{in-in} (closed time path) formalism of nonequilibrium quantum field theory \cite{Keldysh:1964ud, Schwinger:1960qe, Calzetta:1986cq, Niegawa:1999pn, Prokopec:2003pj, Prokopec:2004ic}. Then, the Keldysh-Schwinger contour is stretched from $-\infty$ to $\infty$ and back, allowing for the asymptotic particle states to be used approximatively. We shall assume the approximation is reasonable for systems of our interest.

This work presents an alternative approach to finite-temperature effects based on an \emph{in-out} formalism, the $S$-matrix unitarity, and $CPT$ symmetry.  We show that constructing forward diagrams with each propagator wound on a cylindrical surface any number of times is a helpful heuristic tool \cite{*[{We note that a similar idea of cylindrical diagrams has been previously introduced in }] [{ within the context of infrared finiteness in quantum electrodynamics, where, however, no winding of internal lines and no connection to kinetic theory has been considered.}] Frye:2018xjj} and correct statistical factors for the on-shell intermediate states are obtained. Moreover, the $CPT$ and unitarity relations for the equilibrium reaction rate asymmetries are respected, as they are in zero-temperature calculations \cite{Kolb:1979qa, Roulet:1997xa, Bhattacharya:2011sy, Baldes:2014gca, Racker:2018tzw}. These features of our general framework will be demonstrated within the seesaw type-I leptogenesis, where the results are easily comparable to the broad literature on the subject \cite{Garny:2009rv, Garny:2009qn, Hohenegger:2010ofa, Hohenegger:2010zz, Garbrecht:2010sz, Garbrecht:2011aw, Garbrecht:2018mrp, Garbrecht:2013iga, Anisimov:2010aq, Anisimov:2010dk, Beneke:2010wd}.

\section{Thermal effects at the leading perturbative order.}\label{secII}

In this work, the primary focus is to simplify the inclusion of thermal effects at higher orders. However, to introduce a new diagrammatic concept, we first consider a simple leading-order example of the $N_i\rightarrow lH$ decay. 

In the seesaw type-I model Lagrangian, right-handed neutrino singlets $N_i$ couple to the standard model lepton and Higgs doublets via Yukawa interaction
\begin{align}\label{eq1}
\mathcal{L} \supset -\mathcal{Y}_{\alpha i}\bar{N}_i P_L l_{\alpha} H + \mathrm{H.c.}
\end{align}
where $i$ and $\alpha$ label the right-handed neutrino and lepton family, respectively. Here, the only massive particles are the right-handed neutrinos with Majorana masses denoted $M_i$, while the other particles are massless.

How do these interactions affect the $N_i$ number densities in the expanding universe? At the $\mathcal{O}(\mathcal{Y}^2)$ order, the only contribution expected within the \emph{classical} approach comes from the tree-level decays entering the Boltzmann equation as
\begin{alignat}{1}\label{eq2}
\dot{n}_{N_i}+3\mathcal{H}n_{N_i}= &-\mathring \gamma_{N_i\rightarrow lH\vphantom{\bar{l}\bar{H}}}-\mathring \gamma_{N_i\rightarrow \bar{l}\bar{H}}\\
&+\mathring \gamma_{lH\vphantom{\bar{l}\bar{H}}\rightarrow N_i}+\mathring \gamma_{\bar{l}\bar{H}\rightarrow N_i}\nonumber
\end{alignat}
where $\mathcal{H}$ denotes the Hubble constant while thermally averaged decay rates are conventionally written in the form of
\begin{align}\label{eq3}
\mathring\gamma_{N_i\rightarrow lH}=\int\displaylimits_{N_i\rightarrow lH}{\mathring f}_{N_i}\vert\mathring M\vert^2
\end{align}
with abbreviated notation
\begin{align}\label{eq4}
\hskip-1mm\int\displaylimits_{N_i\rightarrow lH} \hskip-2.5mm=\int[d\mathbf{p}_{N_i}][d\mathbf{p}_{l}][d\mathbf{p}_{H}] (2\pi)^4\delta^{(4)}(p_{N_i}-p_l-p_H)
\end{align}
and $[d\mathbf{p}]=d^3\mathbf{p}/((2\pi)^3 2p^0)$. The small circle in Eq. \eqref{eq3} indicates the reaction rate computed in terms of the classical Maxwell-Boltzmann phase-space densities. We assume kinetic equilibrium for all particle species, including the right-handed neutrinos (see Refs. \cite{Garbrecht:2019zaa, Hahn-Woernle:2009jyb, Basboll:2006yx} for a more detailed discussion). When the temperature drops below their masses, a departure from chemical equilibrium parametrized by $\mu_{N_i}$ occurs. The inclusion of general out-of-equilibrium distributions will be discussed in section \ref{secIV}. In Eq. \eqref{eq3}, the zero-temperature amplitude squared, summed over spin, isospin, and lepton flavors, reads
\begin{align}\label{eq5}
\vert\mathring M\vert^2=4p_{N_i}.p_{l}\sum_\alpha\vert\mathcal{Y}_{\alpha i}\vert^2.
\end{align}
\begin{figure}[t!]
\subfloat{\label{fig1a}}
\subfloat{\label{fig1b}}
\subfloat{\label{fig1c}}
\subfloat{\label{fig1cp}}
\centering\includegraphics[scale=1]{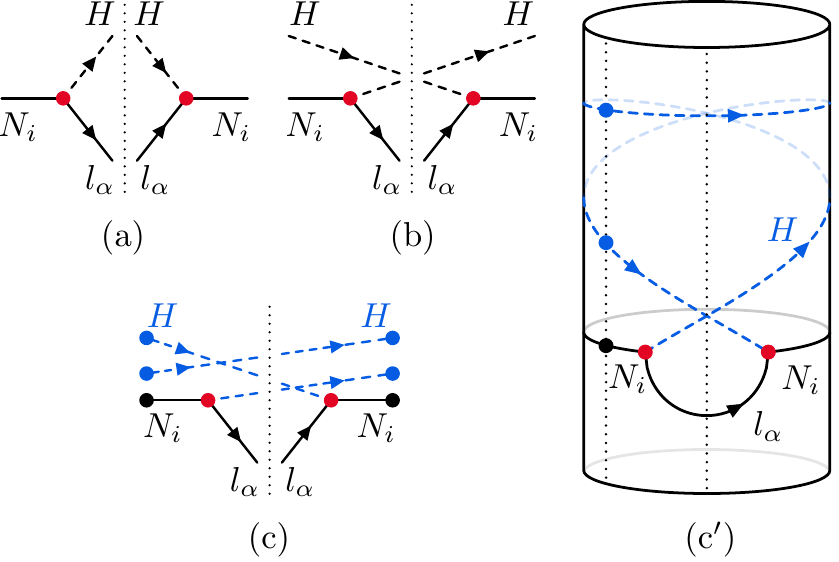}
\caption{\label{fig1} Forward diagrams and their unitary cuts contributing to the neutrino reaction rate in Eq. \eqref{eq2} (Fig. \ref{fig1a} - \ref{fig1c}) and the cylindrical diagram corresponding to Fig. \ref{fig1c} (Fig. \hyperref[fig1cp]{$\mathrm{1c}'$}). Diagrams in Figs. \ref{fig1a}, \ref{fig1b} can also be obtained in the representation of Refs. \cite{Niegawa:1990cd,Niegawa:1997us,Niegawa:1999jr,Ashida:1991ar,Landshoff:1994ud,Wong:2000hq}. The black and blue dots in Figs. \ref{fig1c} and \hyperref[fig1cp]{$\mathrm{1c}'$} do not represent vertices and only serve for the reader's reference.}
\end{figure}

We would like to point out that even when describing particles as classical point-like objects revealing their quantum nature only through the interactions when they come close to each other, the right-hand side of Eq. \eqref{eq2} is incomplete. Let us continue systematically, considering the quantum description of interactions as a black box that can be asked for a complete list of processes occurring at specific order in the couplings. Each reaction contributing to Eq. \eqref{eq2} should be represented by a square of an amplitude, where certain diagrams interfere. As a consequence of $S$-matrix unitarity and optical theorem, each such contribution can be related to cuts of forward diagrams. Examples of those appearing at the second order in the Yukawa coupling, and containing one $N_i$ in the initial state, are shown in Fig. \ref{fig1}. In Fig. \ref{fig1a}, the contribution leading to the simple $N_i$ decay rate in Eq. \eqref{eq3} is shown. Next, in Fig. \ref{fig1b}, the disconnected diagrams involve standalone Higgs lines acting as Lorentz invariant momentum conservation delta-functions, $(2\pi)^3 2p^0 \delta^{(3)}(\mathbf{p}-\mathbf{p}')$, leading to the same amplitude squared as in Fig. \ref{fig1a} though the presence of additional Higgs particle in the initial state brings one extra $\mathring{f}_H$ to the decay rate that is now written as
\begin{align}\label{eq6}
\mathring \gamma_{N_iH\rightarrow lHH}=\int\displaylimits_{N_i\rightarrow lH}{\mathring f}_{N_i}{\mathring f}_H\vert\mathring M\vert^2.
\end{align}

Once the 'Pandora's box' of disconnected diagrams has been opened, one can equally well include any number of Higgses, leptons, and right-handed neutrinos, all as initial states from the black box point of view. However, the disconnected amplitudes should be taken with care. If, for example, the standalone Higgs lines in Fig. \ref{fig1b} were joined to each other and not connected to the rest of the diagram, the contribution to the decay rate would be equal to $\mathring\gamma_{N_i\rightarrow lH}$ from Eq. \eqref{eq3} times the overall number of Higgs particles in the universe and thus already accounted for in Fig. \ref{fig1a}. To avoid such singular and spurious contributions, we should only consider those in which the spectator lines joined together as a single line are connected to the interaction part of the diagram. In other words, we can start drawing forward diagrams such as in Fig. \ref{fig1a} on a cylindrical surface joining the right-handed neutrino lines. Next, we wind the propagators around as shown in Fig. \hyperref[fig1cp]{$\mathrm{1c}'$}. In the end, the contributions of these diagrams will only differ by the powers of phase-space densities and overall sign when the sum of fermionic winding numbers is odd. The sum of all corresponding reaction rates is denoted $\gamma_{N_i\rightarrow lH}$ and equals
\begin{align}\label{eq7}
\int\displaylimits_{N_i\rightarrow lH}f^{\vphantom{\eq}}_{N_i}(1-f^\eq_l)(1+f^\eq_H)\vert\mathring M\vert^2
\end{align}
with newly introduced
\begin{align}\label{eq8}
f_{N_i}=\sum^\infty_{w=1}(-1)^w{\mathring f}_{N_i}^w = \frac{1}{\exp\left\{\frac{E_{N_i}-\mu_{N_i}}{T}\right\}+1}.
\end{align}
Starting the sum with $w=0$, factors $1-f^\eq_l$ and $1+f^\eq_H$ come out on the right-hand side of Eq. \eqref{eq7}. The '$\eq$' superscript denotes the equilibrium Fermi-Dirac or Bose-Einstein distribution, respectively. 

We emphasize that there is no direct manifestation of Pauli's exclusion principle in this picture as we started from the kinetic theory of classical particles. Otherwise, the initial states with identical right-handed neutrinos leading to Eq. \eqref{eq7} could not be considered, and quantum statistics would not emerge. Instead, the alternating sign in Eq. \eqref{eq8} comes from the crossings of legs in Feynman diagrams. This is similar to the $t$- and $u$-channel relative sign in $2\rightarrow 2$ scattering with identical fermions in the final state. 

At this place, the appearance of quantum statistics in Eq. \eqref{eq7} may seem a miraculous coincidence. It will be shown in section \ref{secIV} that it is not so and this appearance is general. Now we continue with thermal corrections to the reaction rate asymmetries, where the inclusion of the proper statistical factors is more subtle.

\section{$CP$ asymmetric thermal rates for leptogenesis.}\label{secIII}

In leptogenesis, the collision terms in the Boltzmann equation for the lepton number generation, to the first order in asymmetric quantities, can be divided into two parts. The \emph{wash-out} terms are proportional to the net number density, i.e. differences between particle and antiparticle densities, and quickly erase any accidentally produced asymmetry if a thermal equilibrium is maintained. The \emph{source} terms, on the other hand, generate the asymmetry when the departure from equilibrium occurs (at temperatures lower than the right-handed neutrino mass) and are expressed in terms of the symmetric parts of the densities and asymmetries of reaction rates.

\begin{figure}[t!]
\subfloat{\label{fig2a}}
\subfloat{\label{fig2b}}
\subfloat{\label{fig2c}}
\centering\includegraphics[scale=1]{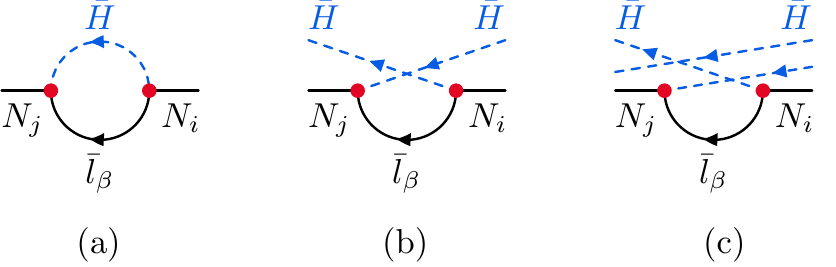}
\caption{\label{fig2} Right-handed neutrino self-energy diagram (Fig. \ref{fig2a}) with thermal corrections to the $\bar{H}$ scalar propagator (Figs. \ref{fig2b} and \ref{fig2c}) modifying its on-shell part.}
\end{figure}

In this section, we focus on thermal corrections to the latter. To that purpose, let us briefly comment on the connection of the presented framework to the common accounting of thermal effects. We focus on the role of the scalar $\bar{H}$ propagator in the right-handed neutrino self-energy. In Fig. \ref{fig2a}, its contribution takes a usual form of 
\begin{align}\label{eq9}
\frac{\iu}{p_{\bar{H}}^2+\iu\epsilon} = \mathrm{P.V.}\frac{\iu}{p_{\bar{H}}^2}+\pi\delta(p_{\bar{H}}^2).
\end{align}
Whenever the $N_j$ self-energy occurs as a subdiagram, depending on the overall kinematics, the positive frequency on-shell part of this propagator may contribute or not. If it does, the diagrams in Figs. \ref{fig2b} and \ref{fig2c} do as well. However, each of them contributes by some number of $\bar{H}$ particles to the initial state of the process and thus by certain power of $\mathring{f}_{\bar{H}}$ to the corresponding reaction rate. The sum of all such diagrams may be represented by a single self-energy diagram with the scalar propagator replaced by
\begin{align}\label{eq10}
\frac{\iu}{p_{\bar{H}}^2+\iu\epsilon}+2\pi\sum^\infty_{w=1} {\mathring{f}_{\bar{H}}}^w \theta(p_{\bar{H}}^0) \delta(p_{\bar{H}}^2)
\end{align}
equivalent to the usual form of the positive frequency part of a thermal propagator. The negative frequencies contribute by the diagrams similar to those in Figs. \ref{fig2b} and \ref{fig2c} with $\bar{H}$ legs flipped to the opposite side of each diagram. In the $N_i$ decay rate, it is the on-shell part, the second term on the right-hand side of Eq. \eqref{eq9}, that contributes to the $CP$ violation at zero temperature. Therefore, to take the thermal corrections to the reaction rate asymmetries into account, we should consider all the windings of the propagators involved in the contributing diagrams.

As a next step, to keep the simplicity of diagrammatic notation, the standard approach to $CP$ asymmetries relying on the Cutkosky rules and imaginary parts of loop integrals needs to be modified. Here we follow our previous work devoted to asymmetries at zero-temperature (see Ref. \cite{Blazek:2021olf}). Following this procedure will allow us later (in Eqs. \eqref{eq19}) to treat the unitary $CP$-violating cuts in the same way (with the same statistical factors) as for the final states.

\begin{figure*}[t!]
\subfloat{\label{fig3a}}
\subfloat{\label{fig3b}}
\subfloat{\label{fig3c}}
\centering\includegraphics[scale=1]{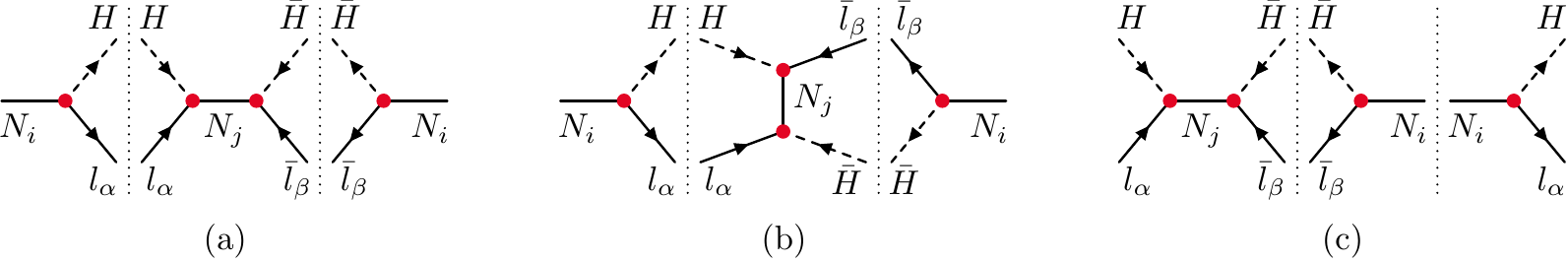}
\caption{\label{fig3} Lepton number violating contributions to the $CP$ asymmetry in the $N_i\rightarrow lH$ decay (Figs. \ref{fig3a} and \ref{fig3b}) and the $s$-channel part of the real-intermediate-state-subtracted $lH\rightarrow\bar{l}\bar{H}$ scattering (Fig. \ref{fig3c}). The latter can be obtained by a cyclic permutation of the diagrams in Fig. \ref{fig3a} using the approach of Refs. \cite{Roulet:1997xa,Blazek:2021olf}.}
\end{figure*}

To sketch the idea briefly, let us expand the unitarity condition $S^\dagger= S^{-1}$ for $S=1+\iu T$.
Employing 
\begin{align}\label{eq11}
\left(1+\iu T\right)^{-1} = 1- \iu T + \iu^2 T^2 - \iu^3 T^3 + \ldots
\end{align}
we obtain
\begin{align}\label{eq12}
\iu T^\dagger = \iu T - \iu^2 T^2 + \iu^3 T^3 -\ldots
\end{align}
where the $T$-matrix elements for the $i\rightarrow f$ reaction can be written as $T_{fi}=(2\pi)^4\delta^{(4)}(p_f-p_i)\mathring{M}_{fi}$. The expansion of Eq. \eqref{eq12} is always truncated at a finite order given by the power of the coupling constants. For the squared amplitude entering the reaction rate, we use
\begin{align}\label{eq13}
\vert T_{fi}\vert^2=-\iu T^{\vphantom{\dagger}}_{if}\iu T^{\vphantom{\dagger}}_{fi}
+\sum_n\iu T^{\vphantom{\dagger}}_{in}\iu T^{\vphantom{\dagger}}_{nf}\iu T^{\vphantom{\dagger}}_{fi}-\ldots
\end{align}
that should be divided by $V_4=(2\pi)^4\delta^{(4)}(0)$ to obtain $(2\pi)^4\delta^{(4)}(p_f-p_i)\vert\mathring{M}_{fi}\vert^2$. Each term in Eq. \eqref{eq13} may be interpreted as a contribution of a forward diagram with one or more simultaneous on-shell cuts. With $\iu T^\dagger$ from Eq. \eqref{eq12}, the $CP$ asymmetry $\Delta \vert T^{\vphantom{\dagger}}_{fi}\vert^2 = \vert T^{\vphantom{\dagger}}_{fi}\vert^2 - \vert T^{\vphantom{\dagger}}_{if}\vert^2$ can be expressed as \cite{Blazek:2021olf}
\begin{align}\label{eq14}
\Delta \vert T^{\vphantom{\dagger}}_{fi}\vert^2&=\sum_{n}(\iu T_{in} \iu T_{nf} \iu T_{fi} - \iu T_{if} \iu T_{fn} \iu T_{ni})\\
&-\sum_{n,m}(\iu T_{in} \iu T_{nm} \iu T_{mf} \iu T_{fi} - \iu T_{if} \iu T_{fm} \iu T_{mn} \iu T_{ni})\nonumber\\
&+\vphantom{\sum_{n}}\ldots\nonumber
\end{align}
From Eq. \eqref{eq14}, the $CPT$ and unitarity relations
\begin{align}\label{eq15}
\sum_f \Delta \vert T^{\vphantom{\dagger}}_{fi}\vert^2 = 0
\end{align}
are manifest. Analogous relations hold for the equilibrium reaction rate asymmetries when the classical Boltzmann approach is used \cite{Kolb:1979qa,Roulet:1997xa,Bhattacharya:2011sy,Baldes:2014gca,Racker:2018tzw}. Concerning thermal-corrected rates, similar cancelations were studied in Ref. \cite{Hook:2011tk} in the scenario of asymmetric dark matter freeze-in.

The meaning of Eq. \eqref{eq14} can be understood from Figs. \ref{fig3a} and \ref{fig3b}. In both diagrams, $i = \vert N_i\rangle$, $f=\vert l_\alpha H\rangle$ and $n =\vert \bar{l}_\beta\bar{H}\rangle$. After subtracting similar contributions with the intermediate states in a reversed order (by means of Eq. \eqref{eq14}), the asymmetry of the squared amplitude for the $N_i\rightarrow lH$ decay, before the intermediate state momentum integration is performed, reads
\begin{align}\label{eq16}
\Delta \vert\mathring M\vert^2=&4p_{l\vphantom{\bar{l}}}.p_{\bar{l}}\sum_{j\neq i} 2\Im\left[\left(\mathcal{Y}\mathcal{Y}^\dagger\right)_{ij}\hskip-1mm\hphantom{}^2\right]\\
&\times\left(\frac{2M_i M_j}{M^2_i-M^2_j}+\frac{M_i M_j}{(p_{l\vphantom{\bar{l}}}-p_{\bar{H}})^2-M^2_j}\right).\nonumber
\end{align}
Here the two terms correspond to the diagrams in Figs. \ref{fig3a} and \ref{fig3b}, respectively. Within the standard Cutkosky approach, the same contributions to the asymmetry are obtained from the imaginary part of the self-energy and vertex loop integrals. Integrating Eq. \eqref{eq16} over the $l$, $H$, $\bar{l}$ and $\bar{H}$ phase-space, and dividing by twice the symmetric part of the squared amplitude from Eq. \eqref{eq5}, we obtain the usual expression for the lepton number violating part of the zero-temperature $CP$ violating parameter \cite{Covi:1996wh}
\begin{align}\label{eq17}
\epsilon_i=\frac{1}{8\pi}\sum_{j\neq i}\Im\left[\left(\mathcal{Y}\mathcal{Y}^\dagger\right)_{ij} \hskip-1mm\hphantom{}^2\right] g(M^2_j/M^2_i)
\end{align}
where
\begin{align}\label{eq18}
g(x)=\frac{\sqrt{x}}{1-x}+\sqrt{x}\left[1-(1+x)\ln\left(1+\frac{1}{x}\right)\right].
\end{align}
Moreover, the asymmetries of the  $\bar{l}\bar{H}\rightarrow N_i$ inverse decay and $lH\rightarrow \bar{l}\bar{H}$ scattering (see Fig. \ref{fig3c}), summed over spin, isospin, and flavor, can be obtained by cyclic permutations of the diagrams in Figs. \ref{fig3a} and \ref{fig3b}, leading to the same $\Delta \vert\mathring M\vert^2$ as in Eq. \eqref{eq16}. It is the advantage of Eq. \eqref{eq14} that it allows treating the asymmetry in scattering, at this order known as the real-intermediate-state-subtracted scattering, in the same way as the asymmetry in the $N_i$ decay. In the standard approach, the tree-level $lH\rightarrow \bar{l}\bar{H}$ amplitude is considered with finite right-handed neutrino width \cite{Kolb:1979qa}. Taking the square and subtracting the part quadratic in the width results in the same asymmetry in $lH\rightarrow \bar{l}\bar{H}$ as can be obtained directly using Eq. \eqref{eq14} and the diagram from Fig. \ref{fig3c} with its $t$-channel counterpart (see Ref. \cite{Blazek:2021olf} for the details). This feature is crucial for the inclusion of thermal corrections to the reaction rate asymmetries entering the lepton number source term. Those are obtained by including windings of the $l$, $H$, $\bar{l}$, $\bar{H}$, and $N_i$ lines in Fig. \ref{fig3}, summing over all winding numbers, the powers of circled densities. Then for the asymmetry in $N_i\rightarrow lH$, the $\bar{l}$, $\bar{H}$ intermediate state receives the same quantum statistical factors as were derived in section \ref{secII} for the $l$, $H$ final states.
\begin{widetext}
\begin{subequations}\label{eq19}
\begin{alignat}{1}
\Delta\gamma_{N_i\rightarrow lH}=&
\int\displaylimits_{N_i\rightarrow lH\rightarrow\bar{l}\bar{H}}
f^{\vphantom{\eq}}_{N_i\vphantom{\bar{N}}}\left(1-f^\eq_{l\vphantom{\bar{l}}}\right)\left(1+f^\eq_{H\vphantom{\bar{H}}}\right)\left(1-f^\eq_{\bar{l}}\right) \left(1+f^\eq_{\bar{H}}\right)
\Delta \vert\mathring M\vert^2,\label{eq19a}\\
\Delta\gamma_{\bar{l}\bar{H}\rightarrow N_i}=&
\int\displaylimits_{N_i\rightarrow lH\rightarrow\bar{l}\bar{H}}
f^\eq_{\bar{l}}f^\eq_{\bar{H}} \big(1-f^{\vphantom{\eq}}_{N_i\vphantom{\bar{N}}}\big) \left(1-f^\eq_{l\vphantom{\bar{l}}}\right) \left(1+f^\eq_{H\vphantom{\bar{H}}}\right)
\Delta \vert\mathring M\vert^2,\label{eq19b}\\
\Delta\gamma_{lH\rightarrow\bar{l}\bar{H}}=&
\int\displaylimits_{N_i\rightarrow lH\rightarrow\bar{l}\bar{H}}
f^\eq_{l\vphantom{\bar{l}}}f^\eq_{H\vphantom{\bar{H}}} \left(1-f^\eq_{\bar{l}}\right) \left(1+f^\eq_{\bar{H}}\right) \big(1-f^{\vphantom{\eq}}_{N_i\vphantom{\bar{N}}}\big)
\Delta \vert\mathring M\vert^2.\label{eq19c}
\end{alignat}
\end{subequations}
\end{widetext}
Here the phase-space integration is understood as in Eq. \eqref{eq4} with additional $(2\pi)^4\delta^{(4)}(p_{l\vphantom{\bar{l}}}+p_{H\vphantom{\bar{H}}}-p_{\bar{l}}-p_{\bar{H}})$ and the $\bar{l}$, $\bar{H}$ three-momentum integrations. We further parametrize $f^{\vphantom{\eq}}_{N_i}$ as $f^{\vphantom{\eq}}_{N_i}=f^\eq_{N_i}+\delta f^{\vphantom{\eq}}_{N_i}$. Note that again, we use the circled notation in Eq. \eqref{eq19} to emphasize the calculation in terms of the zero-temperature Feynman rules. One of this paper's points is that no 'uncircled' amplitudes are needed to include quantum thermal corrections to the Boltzmann equation. 

When the detailed balance condition in the $lH\leftrightarrow\bar{l}\bar{H}$ scattering is taken into account, one can see that the asymmetries in Eqs. \eqref{eq19b} and \eqref{eq19c} are equal. Moreover, replacing $\bar{l}\bar{H}$ with $lH$ and vice versa in Fig. \ref{fig3} changes the order of the two terms in Eq. \eqref{eq14}, leading to an extra minus sign for $lH\rightarrow N_i$ asymmetry. Thus we obtain an example of $CPT$ and unitarity relations
\begin{align}\label{eq20}
\Delta\gamma_{lH\vphantom{\bar{lH}}\rightarrow N_i}+\Delta\gamma_{lH\rightarrow\bar{l}\bar{H}}=0
\end{align}
that is well known in the literature \cite{Kolb:1979qa,Roulet:1997xa,Bhattacharya:2011sy,Baldes:2014gca,Racker:2018tzw}.  Here, however, crucial differences appear, and they belong to the main results of this paper. The reaction rate asymmetries are computed with statistical factors for the final and intermediate states due to the summation of the windings on the cylinder. Thus it is remarkable that employing the cylindrical diagrammatic representation of the previous section allows us to rephrase the thermal-corrected asymmetries in terms of the zero-temperature ones. Relations for thermal-corrected reaction rates have been considered in Ref. \cite{Hook:2011tk}. There the standard Cutkosky approach is applied to asymmetry freeze-in in two sectors at different temperatures. The point of the calculation is the extraction of the loop's imaginary part. Sticking to these thermal cutting rules becomes rather complicated when higher-order terms of the expansion in Eq. \eqref{eq14} are needed (see Ref. \cite{Blazek:2021olf} for the zero-temperature case). Another advantage of Eq. \eqref{eq14} over the common procedure involving the loop's imaginary part extraction is that the unitarity relations, such as in Eq. \eqref{eq20}, are always evident and simple to check.

The lepton number asymmetry evolution arises from the difference of the Boltzmann equations for lepton and antilepton number densities. Considering the reactions mentioned above, the complete $\mathcal{O}(\mathcal{Y}^4)$ thermal-corrected source term, earmarked at the beginning of this section, equals to
\begin{align}\label{eq21}
\Delta\gamma_{N_i\rightarrow lH\vphantom{\bar{l}}}-\Delta\gamma_{lH\rightarrow N_i\vphantom{\bar{l}}}-2\Delta\gamma_{lH\rightarrow\bar{l}\bar{H}}.
\end{align}
Using the detailed balance conditions and Eq. \eqref{eq19}, we can rewrite the source term in Eq. \eqref{eq21} as \cite{Garny:2009rv,Garny:2009qn,Garny:2010nj}
\begin{widetext}
\begin{align}\label{eq22}
\Delta\gamma_{N_i\rightarrow lH\vphantom{\bar{l}}}-\Delta\gamma_{lH\rightarrow\bar{l}\bar{H}}=
\int\displaylimits_{N_i\rightarrow lH\rightarrow\bar{l}\bar{H}}
\delta f^{\vphantom{\eq}}_{N_i}\big(1-f^\eq_{l\vphantom{\bar{l}}}\big) \big(1+f^\eq_{H\vphantom{\bar{H}}}\big) \big(1-f^\eq_{\bar{l}}+f^\eq_{\bar{H}}\big)
\Delta \vert\mathring M\vert^2.
\end{align}
\end{widetext}
Interestingly enough, the last bracket in Eq. \eqref{eq22} is linear in $\bar{l}$, $\bar{H}$ distributions. Refs. \cite{Garny:2009rv,Garny:2009qn} pointed out that using a finite-temperature generalization of Cutkosky cutting rules \cite{Kobes:1985kc,Kobes:1986za} leads to the wrong statistical factors with incorrect additional $-2f^\eq_{\bar{l}}f^\eq_{\bar{H}}$ term \cite{Covi:1997dr,Giudice:2003jh,Davidson:2008bu}. Instead, the causal $n$-point functions \cite{Garny:2010nj} or direct closed-time-path derivation should be used \cite{Garny:2009rv,Garny:2009qn,Hohenegger:2010ofa,Hohenegger:2010zz,Garbrecht:2010sz,Garbrecht:2011aw,Garbrecht:2018mrp,Garbrecht:2013iga,Anisimov:2010aq,Anisimov:2010dk,Beneke:2010wd}. Remarkably, our result is in exact agreement with the latter, which seems to be the first time the correct form of Eq. \eqref{eq22} is derived using the Boltzmann equation and $S$-matrix elements \footnote{See the discussion at the end of Section 6 in Ref. \cite{Garbrecht:2013iga}}.

\section{Quantum kinetic theory and general one-particle distributions.}\label{secIV}

Up until now, it might have seemed that the presented method works for equilibrium densities only. Here we show that it is suitable for nonequilibrium cases too, and that it is not an accident. In general, a quantum state of a many-particle system is given in terms of a density matrix $\hat{\rho}$. Its hermiticity and positive definiteness allows us to write
\begin{align}\label{eq23}
\hat{\rho}=\frac{1}{\mathcal{Z}}\exp\left\{-\hat{\mathcal{F}}\right\}, \mathcal{Z}=\mathrm{Tr}\exp\left\{-\hat{\mathcal{F}}\right\}
\end{align}
with a hermitian operator $\hat{\mathcal{F}}$ \cite{PhysRevB.44.6104}. To simplify the description, a finite volume $V_3$ and time $T_1$ are considered. The one-particle states are labeled simply by $p, q, \ldots$ suppressing the information on particle species and their degrees of freedom. We proceed as if all particles were bosonic \footnote{For fermions, the derivation is analogous to the bosonic case. However, the only possible nonzero factors from Wick contractions, such as in Eq. \eqref{eq29}, equal to $1$. The cylindrical representation then comes from expansion $1/\mathcal{Z}_r= 1/(1+\exp\{-\mathcal{F}_r\})$ as a geometric series.}. In the Boltzmann equation, we only deal with one-particle densities. Therefore, we approximate operator $\hat{\mathcal{F}}$ by
\begin{align}\label{eq24} 
\hat{\mathcal{F}}=\sum_p\mathcal{F}^{\vphantom{\dagger}}_p a^\dagger_p a^{\vphantom{\dagger}}_p \quad\text{with}\quad[a^{\vphantom{\dagger}}_p, a^\dagger_q]=\delta^{\vphantom{\dagger}}_{pq}.
\end{align}
Tracing over the Fock space basis $\vert i_1, i_2, \ldots\rangle$ labeled by one-particle occupation numbers, we obtain  \cite{*[{The approach presented here is a straightforward generalization of what can be found in }] [{ for the equilibrium case.}] Landshoff:1998ku}
\begin{align}\label{eq25}
\mathcal{Z}=\sum_{\{i\}}\exp\left\{-\mathcal{F}_1 i_1 -\mathcal{F}_2 i_2 -\ldots\right\}
=\prod_p \mathcal{Z}_p
\end{align}
with $\mathcal{Z}_p=\exp\{\mathcal{F}_p\}/(\exp\{\mathcal{F}_p\}-1)$. The mean occupation number of a one-particle state equals
\begin{align}\label{eq26}
f_p=\mathrm{Tr}\left[\hat{\rho}a^\dagger_p a^{\vphantom{\dagger}}_p\right]=\frac{1}{\exp\{\mathcal{F}_p\}-1}.
\end{align}
In analogy to the equilibrium case, it is natural to introduce $\mathring{f}_p = \exp\left\{-\mathcal{F}_p\right\}$. In practice, $\mathring{f}_p$ is determined in terms of $f_p$ through Eq. \eqref{eq26}. 

Using $\hat{\rho}' = S\hat{\rho}S^\dagger$ to describe the temporal evolution, the $S$-matrix unitarity leads to \cite{McKellar:1992ja}
\begin{align}\label{eq27}
\hat{\rho}' -\hat{\rho} = T\hat{\rho}T^\dagger -\frac{1}{2}TT^\dagger\hat{\rho} -\frac{1}{2}\hat{\rho}TT^\dagger+\ldots
\end{align}
where $\ldots$ stands for the terms that vanish in  $\mathrm{Tr}[\hat{\rho}a^\dagger_p a^{\vphantom{\dagger}}_p]$ for $\hat{\mathcal{F}}$ in Eq. \eqref{eq24}. From Eqs. \eqref{eq27} and \eqref{eq12} we obtain
\begin{widetext}
\begin{align}\label{eq28}
f'_p-f^{\vphantom{'}}_p &= \mathrm{Tr}\Big[a^\dagger_p a^{\vphantom{\dagger}}_p\Big(T\hat{\rho}T^\dagger -\hat{\rho}TT^\dagger\Big)\Big] = 
\sum^{\infty}_{k=1}(-1)^k\mathrm{Tr}\Big[a^\dagger_p a^{\vphantom{\dagger}}_p
\Big(\iu T\hat{\rho}(\iu T)^k -\hat{\rho}(\iu T)^{k+1} \Big)\Big]\\
&=\frac{1}{\mathcal{Z}}\sum^{\infty}_{k=1}(-1)^k \sum_{\{i\}}\sum_{\{n\}}
\left(n_p - i_p\right)\mathring{f}^{i_1}_1\mathring{f}^{i_2}_2\ldots (\iu T)^k_{in}\iu T^{\vphantom{k}}_{ni}
\nonumber
\end{align}
\end{widetext}
where $\{i\}$, $\{n\}$ denote sequences of one-particle occupation numbers. Dividing Eq. \eqref{eq28} by $T_1 V_3$ gives a change of the particle density per unit time on the left-hand side, while the reaction rates counting the interactions per unit volume per unit time appear to the right. The $T$-matrices in Eq. \eqref{eq28} are computed employing Wick's contractions represented by Feynman diagrams. The trace allows to draw them as diagrams on a cylindrical surface with two or more on-shell cuts. The two cuts that are always there are marked by $i$ and $n$. For each term in Eq. \eqref{eq28}, the one-particle states other than $p$ fall into two distinct groups. 

In the first group, there are those that do not participate in interactions -- their $a$'s or $a^\dagger$'s are not contracted with fields in any of the $T$-matrices. If $q$ is such a state, at the diagrammatic level, it can be represented by a standalone loop whose winding number is $i_q=n_q$. Therefore, in addition to each diagram containing vertices, there are many contributions from cuts of such standalone loops bringing in some powers of circled densities. Finally, they completely factor out by the contribution of the same states to $\mathcal{Z}$. 

The second group then contains one-particle states connected to a vertex at least in one diagram. Within the chain of $k+1$ diagrams in Eq. \eqref{eq28}, for each state labeled by $r\neq p$, three cases may occur: 1. the occupation number $i_r$ goes up by one unit in some $T$, then down in another, 2. the same in a reversed order, 3. $i_r$ is lowered and increased (the external $a$'s contracted with $a^\dagger_r a^{\vphantom{\dagger}}_r$ from the fields) in a single $T$. At the level of Wick's contractions, the first case leads to
\begin{align}\label{eq29}
\left\langle 0 \left\vert
\frac{1}{\sqrt{n_r!}}a_r^{n_r} a^\dagger_r \frac{1}{\sqrt{i_r!}} a^{\dagger i_r}_r
\right\vert 0 \right\rangle \quad\rightarrow\quad \sqrt{1+i_r} 
\end{align}
in a $T$-matrix with $n_r=1+i_r$. The opposite process, with $a^\dagger_r$ after $a_r^{n_r}$ replaced by $a^{\vphantom{\dagger}}_r$ in Eq. \eqref{eq29}, brings in another factor of $\sqrt{1+i_r}$. Thus, in this case, we get
\begin{align}\label{eq30}
\sum^\infty_{i_r=0}(1+i_r)(\mathring{f}_r)^{i_r} = \mathcal{Z}_r \sum^\infty_{i_r=0}(\mathring{f}_r)^{i_r} = \mathcal{Z}_r (1+f_r)
\end{align}
where $\mathcal{Z}_r$ is canceled by $\mathcal{Z}$ in Eq. \eqref{eq28}. Similarly, in the second and third case
\begin{align}\label{eq31}
\sum^\infty_{i_r=1}i_r(\mathring{f}_r)^{i_r} = \mathcal{Z}_r \sum^\infty_{i_r=1}(\mathring{f}_r)^{i_r}= \mathcal{Z}_r f_r
\end{align}
is obtained. Note that we do not consider more contractions within the external states and fields. For $V_3\rightarrow\infty$, such terms can be neglected as they are restricted to a lower-dimensional submanifold of the phase-space.

Finally, state $p$ in Eq. \eqref{eq28} requires $n_p\neq i_p$ such that one of the above three cases, with $n_p-i_p=1$ (case 1.) or $-1$ (cases 2. and 3.), takes place in $T_{ni}$. As in Eqs. \eqref{eq30} and \eqref{eq31}, we obtain factors $\mathcal{Z}_p(1+f_p)$ or $-\mathcal{Z}_pf_p$. These correspond to the gain or loss terms, respectively, in the Boltzmann equation, while the presence of $\mathcal{Z}_p$ ensures that $1/\mathcal{Z}$ in Eq. \eqref{eq28} is completely canceled. Moreover, the sums in Eqs. \eqref{eq30} and \eqref{eq31} \emph{imply that occupation numbers higher than one may be represented by windings of the external lines} in exact agreement with what we saw in sections \ref{secII} and \ref{secIII}.

\section{Conclusions.}

A diagrammatic concept connecting the classical Boltzmann equation and quantum kinetic theory has been introduced in this work. Statistical factors due to on-shell intermediate states have been formally represented by cuts of forward diagrams with multiple spectator lines. The resulting reaction rates and their asymmetries have been expressed in terms of zero-temperature amplitudes, while the statistical factors arose from explicit diagrammatic summation over propagator windings on a cylinder. We showed how this relation is connected to the quantum formulation of kinetic theory and that it is not an accident. Furthermore, for the first time the unitarity and $CPT$ relations between thermal-corrected reaction rate asymmetries have been formulated analogously to the zero-temperature case -- see Eq. \eqref{eq20} for the simplest example -- and less trivial examples will be considered in our follow-up studies.

\begin{acknowledgements}
We thank our colleague, Vladim\'ir Balek, for reading the manuscript and to Fedor \v{S}imkovic for his long-term support. The authors were supported by the Slovak Ministry of Education Contract No. 0243/2021.
\end{acknowledgements}

\bibliographystyle{apsrev4-1.bst}
\bibliography{CLANOK.bib}

\end{document}